\documentclass[12pt]{article}
\usepackage{epsf,latexsym,rotate}
\usepackage[ps,arc,frame]{xy}
\usepackage{amsfonts,amssymb} 
\epsfverbosetrue
\usepackage[ps,arc,frame]{xy}
\usepackage{graphicx}
\usepackage{slashed}

\newcommand{\lapprox}{%
\mathrel{%
\setbox0=\hbox{$<$}
\raise0.6ex\copy0\kern-\wd0
\lower0.65ex\hbox{$\sim$}
}}
\newcommand{\gapprox}{%
\mathrel{%
\setbox0=\hbox{$>$}
\raise0.6ex\copy0\kern-\wd0
\lower0.65ex\hbox{$\sim$}
}}

\newcommand{\double}[1]{\mathbb{#1}}

\newcommand{\cc}{\double{C}}

\newcommand{\aaa}{\mathcal{A}}
\newcommand{\ccc}{\mathcal{C}}

\newcommand{\hhh}{\double{H}}
\newcommand{\mm}{\mathcal{M}}

\newcommand{\pp}{\pmatrix}

\newcommand{\dd}{\mathcal{D}}

\newcommand{\de}{\hbox{\rm{d}}}

\newcommand{\pa}{\partial}

\newcommand{\ul}[1]{\underline{#1}}
\newcommand{\ot}{\otimes}
\newcommand{\op}{\oplus}

\newcommand{\bb}{\begin{eqnarray}}
\newcommand{\ee}{\end{eqnarray}}
\newcommand{\eee}{\nonumber\end{eqnarray}}
\newcommand{\qq}{\quad}

\newcommand{\rxyh}[2]{{\begin{xy} 0;<2mm,0mm>:<0mm,2mm>::0;0,
,(5,-2)*{a}
,(10,-2)*{b}
,(15,-1.8)*{\bar{b}}
,(20,-2)*{c}
,(25,-1.8)*{d}
,(30,-1.8)*{\bar{d}}
,(2,-5)*{a}
,(2,-10)*{b}
,(1.8,-15)*{\bar{b}}
,(2,-20)*{c}
,(1.8,-25)*{d}
,(1.8,-30)*{\bar{d}}
,(5,-5)*\cir(#1,0){}
,(10,-5)*\cir(#1,0){}
,(15,-5)*\cir(#1,0){}
,(20,-5)*\cir(#1,0){}
,(25,-5)*\cir(#1,0){}
,(30,-5)*\cir(#1,0){}
,(5,-10)*\cir(#1,0){}
,(10,-10)*\cir(#1,0){}
,(15,-10)*\cir(#1,0){}
,(20,-10)*\cir(#1,0){}
,(25,-10)*\cir(#1,0){}
,(30,-10)*\cir(#1,0){}
,(5,-15)*\cir(#1,0){}
,(10,-15)*\cir(#1,0){}
,(15,-15)*\cir(#1,0){}
,(20,-15)*\cir(#1,0){}
,(25,-15)*\cir(#1,0){}
,(30,-15)*\cir(#1,0){}
,(5,-20)*\cir(#1,0){}
,(10,-20)*\cir(#1,0){}
,(15,-20)*\cir(#1,0){}
,(20,-20)*\cir(#1,0){}
,(25,-20)*\cir(#1,0){}
,(30,-20)*\cir(#1,0){}
,(5,-25)*\cir(#1,0){}
,(10,-25)*\cir(#1,0){}
,(15,-25)*\cir(#1,0){}
,(20,-25)*\cir(#1,0){}
,(25,-25)*\cir(#1,0){}
,(30,-25)*\cir(#1,0){}
,(5,-30)*\cir(#1,0){}
,(10,-30)*\cir(#1,0){}
,(15,-30)*\cir(#1,0){}
,(20,-30)*\cir(#1,0){}
,(25,-30)*\cir(#1,0){}
,(30,-30)*\cir(#1,0){}
#2\end{xy}}}

\begin{document}

\font\twelve=cmbx10 at 13pt
\font\eightrm=cmr8

\thispagestyle{empty}

\begin{center}

CENTRE DE PHYSIQUE TH\'EORIQUE $^1$ \\ CNRS--Luminy, Case
907\\ 13288 Marseille Cedex 9\\ FRANCE\\

\vspace{2cm}

{\Large\textbf{Seesaw and noncommutative geometry}} \\

\vspace{1.5cm}

{\large Jan-H. Jureit $^2$, Thomas Krajewski $^3$,\\ Thomas
Sch\"ucker $^4$, Christoph A. Stephan $^5$}

\vspace{2cm}

{\large\textbf{Abstract}}
\end{center}
The 1-loop corrections to the seesaw mechanism in the noncommutative standard model are computed. Other consequences of the Lorentzian signature in the inner space are summarised. \vspace{1.2cm}

${}$\hfil {\it dedicated to Alain Connes on the occasion of his 60th  birthday}\break
\vskip 1,0 truecm

\noindent
PACS-92: 11.15 Gauge field theories\\
MSC-91: 81T13 Yang-Mills and other gauge theories

\vskip 1truecm

\noindent 0705.0489\\

\vspace{1.5cm}
\noindent $^1$ Unit\'e Mixte de Recherche  (UMR 6207)
du CNRS  et des Universit\'es Aix--Marseille 1 et 2 et  Sud
Toulon--Var, Laboratoire affili\'e \`a la FRUMAM (FR 2291)\\
$^2$  also at Universit\"at Kiel, jureit@cpt.univ-mrs.fr\\
$^3$  also at Universit\'e Aix--Marseille 1,
krajew@cpt.univ-mrs.fr\\
$^4$ also at Universit\'e Aix--Marseille 1,
thomas.schucker@gmail.com \\
$^5$ also at Universit\'e Aix--Marseille 1,
christophstephan@gmx.de\\

Understanding the origin of the standard model is currently one of
most challenging issues in high energy physics. Indeed, despite its
experimental successes, it is fair to say that its structure remains
a mystery. Moreover, a better understanding of its structure would
provide us with a precious clue towards its possible extensions.

This can be achieved in the framework of noncommutative geometry
\cite{book}, which is a branch of mathematics pioneered by Alain Connes,
 aiming at a generalisation of geometrical ideas to spaces whose
coordinates fail to commute. The physical interpretation of the spectral action
principle \cite{cc} and its confrontation with present-day experiment still
require some contact with the low energy physics. This follows from
the standard Wilsonian renormalization group idea. The spectral
action provides us with a bare action supposed to be valid at a
very high energy of the order of the unification scale. Then,
evolving down to the electroweak scale yields the effective low
energy physics. This line of thought is very similar to the one
adopted in grand unified theories. Indeed, in a certain sense
models based on non commutative geometry can be considered as
alternatives to grand unification that do not imply proton decay.

Ten years after its discovery \cite{cc}, the spectral action has  recently received new impetus \cite{c06,barr,mc2} by allowing a  Lorentzian signature in the internal space. This mild modification  has three consequences: (i)
the fermion-doubling problem \cite{2f} is solved elegantly, (ii)
Majorana masses and consequently the popular seesaw mechanism are  allowed for, (iii) the Majorana masses in turn decouple the Planck mass from the $W$ mass.
Furthermore, Chamseddine, Connes \& Marcolli point out an additional  constraint on the coupling constants tying the sum of all Yukawa  couplings squared to the weak gauge coupling squared. This relation  already holds for Euclidean internal spaces
\cite{thum}.

\section{Constraints on gauge groups and representations}

There are two ways to extract the gauge group $G$ from the spectral  triple.
The first way defines the gauge group to be the unimodular (i.e. of  unit determinant) unitary group \cite{real,mc2} of the associative  algebra, which by the faithful representation immediately acts on the  Hilbert space.
The second way follows general relativity whose invariance group is  the group of diffeomorphisms of $M$ (general coordinate  transformations). In the algebraic formulation this is the group of  algebra automorphisms. Indeed Aut$(\ccc^\infty(M))={\rm Diff}(M).$ We  still have to lift the diffeomorphisms to the Hilbert space. This  lift is double-valued and its image is the semi-direct product of the  diffeomorphism group with the local spin group \cite{lift}.
Up to possible central $U(1)$s and their central charges, the two  approaches coincide.

There are other constraints on the fermionic representations coming
 from the axioms of the spectral triple. They are conveniently
captured in Krajewski diagrams which classify all possible finite  dimensional spectral triples \cite{kps}. They do for spectral triples  what the Dynkin and weight diagrams do for groups and  representations. Figure 1 shows the Krajewski diagram of the standard  model in Lorentzian signature with one generation of fermions  including a right-handed neutrino.

\begin{center}
\begin{tabular}{c}
\rxyh{0.7}{
,(5,-20)*\cir(0.3,0){}*\frm{*}
,(5,-25)*\cir(0.3,0){}*\frm{*}
,(5,-20);(10,-20)**\dir{-}?(.4)*\dir{<}
,(5,-20);(15,-20)**\crv{(10,-17)}?(.4)*\dir{<}
,(5,-25);(15,-25)**\crv{(10,-28)}?(.4)*\dir{<}
,(5,-25);(30,-25)**\crv{~*=<2pt>{.}(17.5,-20)}?(.45)*\dir{<}
,(20,-5)*\cir(0.3,0){}*\frm{*}
,(25,-5)*\cir(0.3,0){}*\frm{*}
,(20,-5);(20,-10)**\dir{-}?(.6)*\dir{>}
,(20,-5);(20,-15)**\crv{(17,-10)}?(.6)*\dir{>}
,(25,-5);(25,-15)**\crv{(28,-10)}?(.6)*\dir{>}
,(25,-5);(25,-30)**\crv{~*=<2pt>{.}(20,-17.5)}?(.6)*\dir{>}
,(30,-25)*\cir(0.3,0){}*\frm{*}
,(25,-30)*\cir(0.3,0){}*\frm{*}
,(25,-30);(30,-25)**\dir{--}?(.4)*\dir{<}
} \\ \\
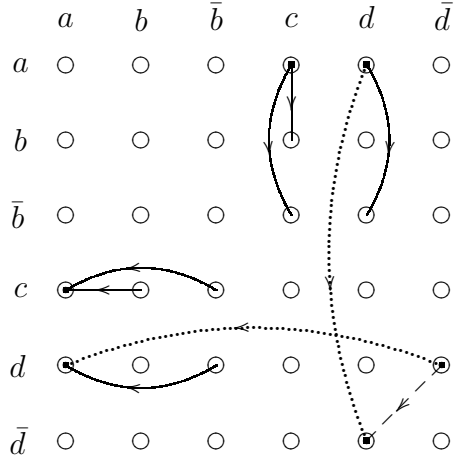
Figure 1: Krajewski diagram of the standard model with right-handed  neutrino \\
\hskip-1.1cm and Majorana-mass term depicted by the dashed arrow.
\\
\end{tabular}
\end{center}

Certainly the most restrictive constraint on the discrete parameters  concerns the scalar representation. In the Yang-Mills-Higgs ansatz it  is an arbitrary input. In the almost commutative setting it is {\it  computed} from the data of the inner spectral triple.

Here we present the inner triple of the standard model with one generation of  fermions including a right-handed neutrino. The algebra has four  summands:
$\aaa=\hhh\op\cc\op M_3(\cc)\op\cc\owns (a,b,c,d),$
the Hilbert space is 32-dimensional
and carries the faithful repesentation
$ \rho(a,b,c,d):=
\rho_{L}\oplus\rho_{R}\oplus{\bar\rho^c_{L}}\oplus{\bar\rho^c_{R}}$
with
\bb\rho_{L}(a):=
a\ot 1_3\oplus a,\ 
\rho_{R}(b,d):=b  1_3\oplus\bar b  1_3\oplus b\oplus d, \nonumber\\[1mm] 
\rho^c_{L}(c,d):=
1_2\ot c\oplus \bar d1_2,\ 
\rho^c_{R}(c,d) :=
c\oplus c\oplus \bar d\oplus\bar d.  
\ee
The Dirac operator reads
\bb  \dd=\pp{0&\mm&0&0\cr
\mm^*&0&0&S\cr
0&0&0&\bar\mm\cr
0&S^*&\bar\mm^*&0},\qq
S=\pp{M_M&0\cr0&0}\ee
where $S$ contains the Majorana mass for the right-handed neutrino and $\mm$
contains the Dirac masses
\bb\mm=
\pp{M_u&0\cr 0&M_d} \ot 1_3
\oplus\pp{M_\nu&0\cr 0&M_e}.\ee
This model is conform with the standard formulation of the axiom of Poincar\'e
duality as stated in \cite{book}. It seems to be closely related to the
older bi-module approach of the Connes-Lott model \cite{lott} which
also exhibits two copies of the complex numbers in the algebra.
In the original version of the standard model with right-handed Majorana
neutrinos Chamseddine, Connes \& Marcoli \cite{mc2} used an
alternative spectral triple based on the algebra $\aaa=\hhh\op\cc\op M_3(\cc)$.
But this spectral triple requires a
subtle change in the formulation of the Poincar\'e duality, i.e. it needs
two elements to generate $KO$-homology as a module over $K_0$.
It should however be noted that right-handed neutrinos that allow for
Majorana-masses always fail to fulfil the axiom of orientability \cite{ko6}
since the representation of the algebra does not allow to construct
a Hochschild-cycle reproducing the chirality operator. For this reason we have drawn the arrows connected to the right-handed neutrinos with broken lines in the Krajewski diagram.

\section{Constraints on dimensionless couplings}

The spectral action counts the number of eigenvalues of the Dirac  operator whose absolute values are less than the energy cut-off $ \Lambda$. On the input side its continuous parameters are: this cut-off, three positive parameters in the cut-off functions and the  parameters of the inner Dirac operator, i.e. fermion masses and  mixing angles. On the output side we have: the cosmological constant,  Newton's constant, the gauge and the Higgs couplings. Therefore there  are constraints, which for the standard model with $N$ generations  and three colours read:
\bb \,\frac{5}{3}\,g_1^2=  g_2^2=g_3^2=\,\frac{3}{N}\,\frac{Y_2^2}{H} \,\frac{\lambda}{24}\,= \,\frac{3}{4N}\,Y_2\,.\label{4con}\ee
Here $Y_2$ is the sum of all Yukawa couplings $g_f$ squared, $H$ is  the sum of all Yukawa couplings raised to the fourth power. Our  normalisations are: $m_f=\sqrt{2}\,(g_f/g_2)\,m_W,$ $(1/2)\,(\pa  \varphi)^2+(\lambda/24)\,\varphi^4$. If we define the gauge group by  lifted automorphisms rather than unimodular unitairies, then we get  an ambiguity parameterized by the central charges. This ambiguity  leaves the $U(1)$ coupling $g_1$ unconstrained and therefore kills  the first of the four constraints (\ref{4con}).

Note that the noncommutative constraints (\ref{4con}) are different
from Veltman's condition \cite{velt}, which in our normalisation reads:
$  {\textstyle\frac{3}{4}}g_2^2+{\textstyle\frac{1}{4}}g_1^2 + {\textstyle\frac{1}{3}}\lambda-2g_t^2=0.$

Of course the constraints (\ref{4con}) are not stable under the  renormalisation group flow and as in grand unified theories we can  only interpret them at an extremely high unification energy $\Lambda $. But in order to compute the evolution of the couplings between our  energies and $\Lambda$ we must resort to the daring hypothesis of the  big desert, popular since grand unification. It says: above presently  explored energies and up to $\Lambda$ no more new particle, no more  new forces exist, with the exception of the Higgs, and that all  couplings evolve without leaving the perturbative domain. In  particular the Higgs self-coupling $\lambda$ must remain positif. In  grand unified theories one believes that new particles exist with  masses of the order of $\Lambda$, the leptoquarks. They mediate  proton decay and stabilize the constraints between the gauge  couplings by a bigger group. In the noncommutative approach we  believe that at the energy $\Lambda$ the noncommutative character of  space-time ceases to be negligible. The ensuing uncertainty relation  in space-time might cure the short distance divergencies and thereby  stabilize the constraints. Indeed Grosse \& Wulkenhaar have an  example of a scalar field theory on a noncommutative space-time whose  $\beta$-function vanishes to all orders \cite{raimar}.

Let us now use the one-loop $\beta$-functions of the standard model   with $N=3$ generations to evolve the constraints (\ref{4con}) from $E= \Lambda$  down to our energies $E=m_Z$. We set:
$ t:=\ln (E/m_Z),\qq \de g/\de t=:\beta _g,\qq \kappa :=(4\pi )^{-2}. $ We will neglect all fermion masses below the top mass and also  neglect threshold effects. We admit a Dirac mass $m_D$ for the $\tau$  neutrino induced by spontaneous symmetry breaking and take this mass  of the order of the top mass. We also admit a Majorana mass $m_M$ for  the right-handed $\tau$ neutrino. Since this mass is gauge invariant  it is natural to take it of the order of $\Lambda$. Then we get two  physical masses for the
$\tau$ neutrino: one is tiny, $m_\ell\sim m_D^2/m_M$, the other is  huge, $m_r\sim m_M.$ This is the popular seesaw mechanism. The renormalisation of these masses is well-known \cite {seesawren}. By the Appelquist-Carazzone decoupling theorem  we distinguish two energy domains: $E>m_M$ and $E<m_M$. In the  latter, the Yukawa coupling of the $\tau$ neutrino drops out of the $ \beta$-functions and is replaced by an effective coupling
\bb k=2\,\frac{g_\nu^2}{m_M}\,, \qq{\rm at\ }E=m_M.\ee
At high energies, $E>m_M$, the $\beta$-functions are \cite{mv,jones}:
\bb \beta _{g_i}&=&\kappa b_ig_i^3,\qq b_i=
{\textstyle
\left( \frac{20}{9} N+\frac{1}{6},-\frac{22}{3}+\frac{4}{3} N+\frac{1} {6},
-11+\frac{4}{3} N\right) },
\\ \cr
\beta _t&=&\kappa
\left[ -\sum_i c_i^ug_i^2 +Y_2 +\,\frac{3}{2}\,g_t^2
\,\right] g_t,\qq
\beta _\nu\ =\ \kappa
\left[ -\sum_i c_i^\nu g_i^2 +Y_2 +\,\frac{3}{2}\,g_\nu^2\,
\right] g_\nu,\\
\beta _\lambda &=&\kappa
\left[ \,\frac{9}{4}\,\left( g_1^4+2g_1^2g_2^2+3g_2^4\right)
-\left( 3g_1^2+9g_2^2\right) \lambda
+4Y_2\lambda -12H+4\lambda ^2\right] ,\ee
with
$ c_i^t=\left( {\textstyle\frac{17}{12}},{\textstyle\frac{9}{4}} , 8 \right) ,
\ 
c_i^\nu =\left( {\textstyle\frac{3}{4}},{\textstyle\frac{9}{4}} , 0 \right) ,\ 
Y_2=3g_t^2+g_\nu^2,
\ 
H=3g_t^4+g_\nu^4.
$
At low energies, $E<m_M$,
the $\beta$-functions are the same except that
$Y_2=3g_t^2$, $H=3g_t^4$ and that $\beta_\nu$ is replaced \cite {seesawren} by:
\bb
\beta _k=\kappa
\left[ -3g_2^2+\,\frac{3}{2}\,-\sum_i c_i^\nu g_i^2 +Y_2 +\,\frac{2} {3}\,\lambda+2Y_2\,
\right] k.\ee
We suppose that all couplings (other than $g_\nu$ and $k$) are  continuous  at $E=m_M$, no threshold effects.
The three gauge couplings decouple from the other equations and have  identical evolutions in both energy domains:
\bb g_i(t)=g_{i0}/\sqrt{1-2\kappa b_ig_{i0}^2t}.\ee
The initial conditions are taken from experiment \cite{data}:
$ g_{10}= 0.3575,\ 
g_{20}=0.6514,\ 
g_{30}=1.221.$
In a first run we leave $g_1$ unconstrained. Then the unification  scale $\Lambda $ is the solution of $g_2(\ln (\Lambda /m_Z))=g_3(\ln  (\Lambda /m_Z))$,
\bb \Lambda = m_Z\exp\frac{g_{20}^{-2}-g_{30}^{-2}}{2\kappa (b_2-b_3)} \,=\,1.1\times 10^{17}\  {\rm GeV},\ee
and is independent of the number of generations.

Then we choose $g_\nu=Rg_t$ at $E=\Lambda$ and $m_M$, and solve  numerically the evolution equations for $\lambda,\ g_t,\ g_\nu$ and $k $ with initial conditions at $E=\Lambda$ from the noncommutative  constraints (\ref{4con}):
\bb g_2^2=\,\frac{3+R^2}{3+R^4}\,\frac{\lambda}{24}\,= \,\frac{3+R^2} {4}\,g_t^2\,.\ee
We note that these constraints imply that all couplings remain  perturbative and
at our energies we obtain the pole masses of the Higgs, the top and  the light neutrino:
\bb m_H^2=\,\frac{4}{3}\,\frac{\lambda(m_H) }{g_2(m_Z)^2}\,m_W^2,\qq
m_t=\sqrt{2}\,\frac{g_t(m_t)}{g_2(m_t)}\,m_W,\qq
m_\ell=\,\frac{k(m_Z)}{g_2(m_Z)^2}\,m_W^2.\ee
A few numerical results are collected in table 1.
\begin{table}[h]
\begin{center}
\begin{tabular}{|c|c|c|c|c|c|c|c|c|c|}
\hline
$g_\nu/g_t|_\Lambda$ & 0 & 1.16& 1.16&1.2&1.2&1.3&1.3&1.4&1.4
\\
$m_M$ [GeV]& *&$2\cdot10^{13}$&$10^{14}$&$2\cdot10^{13}$&$10^{14}$&$3 \cdot10^{13}$&$10^{14}$&$3\cdot10^{13}$&$10^{14}$\\
\hline
$m_t$ [GeV]&186.3&173.3&173.6&172.5&172.8&170.5&170.7&168.4&168.6\\
$m_H$ [GeV]&188.4&170.5&170.8&169.7&170.0&167.7&168.0&165.8&166.1\\
$m_\ell$ [ eV]&0&0.29&0.06&0.30&0.06&0.23&0.07&0.25&0.08\\
\hline
\end{tabular}
\end{center}
\caption{The top, Higgs and neutrino masses as a function of $R$ and  of the Majorana mass for the unification scale $\Lambda = 1.1\times  10^{17}$ GeV}
\end{table}

Note that the Higgs mass is not very sensitive to the three input  parameters,  $\Lambda,\ m_M,$ and $ R=g_\nu(\Lambda)/g_t(\Lambda)$ as  long as they vary in a range reproducing senible masses for the top  and the light neutrino, today $m_t=170.9\,\pm2.6$ GeV and $0.05\ {\rm  eV}\ <\ m_\ell\ <\ 0.3$ eV. Then we have for the Higgs mass
\bb m_H= 168.3\pm 2.5\ {\rm GeV}.\ee

In a second run we use the constraint on the Abelian coupling $\,\frac {5}{3}\,g_1^2=  g_2^2$
to compute the unification scale: \bb  \Lambda = m_Z\exp\frac{g_{20}^ {-2}-(3/5)g_{30}^{-2}}{2\kappa (b_2-(3/5)b_1)}\,=\,9.8\times 10^{12} \  {\rm GeV}.\ee

Note that the third constraint $\,\frac{5}{3}\,g_1^2=  g_3^2$ yields  an intermediate unification scale, $\Lambda =2.4\times 10^{14}\  {\rm  GeV}.$ Again we give a few numerical results, table 2.
\begin{table}[h]
\begin{center}
\begin{tabular}{|c|c|c|c|c|c|}
\hline
$g_\nu/g_t|_\Lambda$ & 0 & 0.95& 1&1.1&1.2
\\
\hline
$m_t$ [GeV]&183.4&173.3&172.3&170.3&168.1\\
$m_H$ [GeV]&188.5&174.9&173.8&171.9&170.2\\
$m_\ell$ [ eV]&0&0.53&0.57&0.67&0.77\\
\hline
\end{tabular}
\end{center}
\caption{The top, Higgs and neutrino masses as a function of $R$ for  the Majorana mass and the unification scale $m_M=\Lambda = 9.8\times  10^{12}$ GeV}
\end{table}

Here the upper bound for the light neutrino mass cannot be met  strictly with $m_M<\Lambda$. This does not worry us because that  bound derives from cosmological hypotheses.
Honouring the constraints for all three gauge couplings then yields  the combined range for the Higgs mass,
\bb m_H= 168.3^{+6.6}_{-2.5}\ {\rm GeV}.\ee

\section{Constraints on the dimensionful couplings}
The spectral action also produces constraints between the quadratic  Higgs coupling, the Planck mass, $m_P^2=1/G$,
and the cosmological constant in terms of the cut-off $\Lambda$ and  of the three moments, $f_0, f_2, f_4$, of the cut-off function. A  step function for example has $2f_0=f_2= f_4$.  Trading the quadratic  Higgs coupling for the $W$ mass, these constraints read:
\bb m_W^2=\frac{45}{4\cdot96}\frac{(3m_t^2+m_\nu^2)^2}{3m_t^4+m_\nu^4} \left[\frac{f_2}{f_4}\,\Lambda^2-
\frac{m_\nu^2}{3m_t^2+m_\nu^2}\,m_M^2\right]
\sim \frac{45}{96}\left[\frac{f_2}{f_4}\,\Lambda^2-
\frac{1}{4}\,m_M^2\right],
\ee\bb
m_P^2&=&\frac{1}{3\pi}\left[
\left(96-4\,\frac{(3m_t^2+m_\nu^2)^2}{3m_t^4+m_\nu^4}\right)f_2\Lambda^2
+2\left(2\,\frac{m_\nu^2(3m_t^2+m_\nu^2)}{3m_t^4+m_\nu^4}-1\right) f_4m_M^2\right]\cr\cr
&\sim&
\frac{1}{3\pi}\left[80 f_2\Lambda^2
+2 f_4m_M^2\right],
\\[1mm]
\Lambda_c&\sim&\frac{1}{\pi m_P^2}[(96\cdot 2 f_0-16 f_22/f_4) \Lambda^4+f_4m_M^4+4f_2\Lambda^2m_M^2].\ee
We have taken three generations, i.e. a 96-dimensional inner Hilbert  space, we only kept the Yukawa couplings of the  top quark and of the  $\tau$ neutrino, and one single Majorana mass in the third  generation. The experts still do not agree whether the  renormalisation  group flow of the quadratic Higgs coupling is  logarithmic or quadratic in the energy $E$. Nobody knows how Newton's  and the cosmological constants depend on energy. Therefore we cannot  put the above constraints to test. It is however reassuring that,  thanks to the seesaw mechanism, a $W$ mass much smaller than the  Planck mass is easily obtained. On the other hand it is difficult to  produce a small cosmological constant.

\section{Is the standard model special?}

Despite all constraints, there is still an infinite number of Yang-Mills-Higgs-Dirac-Yukawa models that can be derived from gravity  using almost commutative geometry. The exploration of this special  class is highly non-trivial and starts with Krajewski diagrams. 

At present there are two approaches in this direction.
The first by Chamseddine, Connes \& Marcolli \cite{mc2} starts from a  left-right symmetric algebra. This algebra admits a privileged bi-module which is identical to the fermionic Hilbert space of the  standard model. The algebra of the standard model is a maximal  subalgebra of the left-right symmetric one and the inner Dirac  operator is almost the maximal operator satisfying the axioms of a  spectral triple. The number of colours and the number of generations  remain unexplained in this approach.

The second approach again has nothing to say about the number of  colours and generations. It is a more opportunistic approach and  copies what grand unified theories did in the frame of  Yang-Mills-Higgs-Dirac-Yukawa theories. There, the idea was to cut  down on the number of possible
models with a `shopping' list of requirements: one wants
the gauge group to be simple,
the fermion representation to be irreducible, complex under the gauge group and free of Yang-Mills anomalies, and
the model to contain the standard model.

Coming back to Connes' noncommutative model building kit, we remark  that the spectral triple of the standard model with one generation of  fermions and a massless neutrino is irreducible. It has another  remarkable property concerning its built-in spontaneous symmetry  breaking: it allows a vacuum giving different masses to the two  quarks although they sit in an isospin doublet. Indeed, in the  majority of noncommutative models the spontaneous symmetry breaking  gives degenerate masses to fermions in irreducible {\it group}  representations. 
For years we have been looking for viable noncommutative models other  than the standard model, without success. We therefore started to  scan the Krajewski diagrams with the following shopping list: we want
the spectral triple to be irreducible,
the fermion representation to be complex under the little group in  every irreducible component, and
possible massless fermions to transform trivially under the little  group.
Furthermore we require
the fermion representation to be free of Yang-Mills and mixed  gravitational Yang-Mills anomalies,
and the spectral triple to have no dynamical degeneracy and the colour  group of every kinematical degeneracy to remain unbroken.

The first step is to get the list of irreducible Krajewski diagrams  \cite{class}. In the case of an inner  spectral triple of Euclidean signature, we have no such diagram for a  simple algebra, one diagram for an algebra with two simple summands,  30 diagrams for three summands, 22 diagrams for four summands,  altogether 53 irreducible diagrams for algebras with up to four  simple summands. The situation simplifies when we go to the  Lorentzian signature where we remain with only 7 diagrams for up to  four summands. These numbers are summarized in table 3.

\begin{table}[h]
\begin{center}
\begin{tabular}{|c|c|c|}
\hline\\[-3mm]
\#(summands) & Eulidean & Lorentzian
\\[1ex]
\hline
1&$\ \ 0$&0\\
2&$\ \ 1$&0\\
3& 30&0\\
4& 22&7\\
$\!\!\!\!\!\!\le 4$& 53&7\\
\hline
\end{tabular}
\end{center}
\caption{The number of irreducible Krajewski diagrams for algebras  with up to 4 simple summands and inner spaces with Euclidean and  Lorentzian signatures }
\end{table}

The second step is to scan all models derived from the irreducible  Krajewski diagrams with respect to our shopping list. In both  signatures we remain with the following models:
The standard model with one generation of fermions, an arbitrary  number of colours $C\ge 2$ and a massless neutrino:
${SU(2)\times U(1)\times SU(C)}
\rightarrow
{U(1)\times SU(C)}$.
We also have three  possible submodels with identical fermion content, but with $SU(2)$  replaced by $SO(2)$, no $W$-bosons, or with $SU(C)$ replaced by $SO(C) $ or $USp(C/2)$, $C$ even, less gluons. There is one more possible  model, the elctro-strong model:
${ U(1)\times SU(C)}
\rightarrow
{U(1)\times SU(C)}.$
The fermionic content is ${\ul C}\op {\ul 1}$, one quark and
one charged lepton.  The two electric charges
are arbitrary but vectorlike. The model has no scalar and no symmetry  breaking.

For those of you who think that our shopping list is unreasonably  restrictive already in the frame of Yang-Mills-Higgs-Dirac-Yukawa  models, here is a large class of such models satisfying our shopping  list: Take any group that has complex representations (like $E_6$)  and take any irreducible, complex, unitary representation of this  group. Put the left- and right-handed fermions in two copies of this  representation, choose the Hilbert space for scalars 0-dimensional  and a gauge-invariant mass for all fermions.

\section{Beyond the standard model}
For many years we have been trying to construct models from  noncommutative geometry that go beyond the standard model \cite {beyond} and we failed to come up with anything physical if it was  not to add more generations and right-handed neutrinos to the  standard model.

The noncommutative constraints on the continuous parameters of the  standard model with $N=4$ generations fail to be compatible with the  hypothesis of the big desert \cite{knecht}.

Since a computer program \cite{prog} was written to list the  irreducible Krajewski diagrams for algebras with more than three  summands we do have a genuine extension of the standard model  satisfying all physical desirata. It comes from an algebra with six  summands \cite{chris} and is identical to the standard model with two  additional leptons $A^{--}$ and $C^{++}$ whose electric charge is two  in units of the electron charge. These new leptons  couple neither to  the charged gauge bosons, nor to the Higgs scalar. Their hypercharges  are vector-like, so that they do not contribute to the electroweak  gauge anomalies. Their masses are gauge-invariant and they constitute  viable candidates for cold dark matter \cite{klop}.

Also, by trial and error, two more models could be found recently \cite{colour,vector}.
The first model is based on an algebra with six summands and adds to the
standard model a lepton-like, weakly charged, left-handed doublet and two right-handed
hypercharge singlets. These four particles are each colour-doublets under
a new $SU(2)_c$ colour group. They participate in the
Higgs mechanism and the noncommutative constraints require their masses
to be around $75$ GeV. Since they have a non-Abelian colour group one
expects a macroscopic confinement \cite{okun} with a confinement radius
of $\sim 10^{-7}$ m. Although these particles have electro-magnetic
charge after symmetry breaking it is not yet clear whether they could have
been detected in existing experiments.

The second model adds to the
standard model three generations of vectorlike doublets with weak and
hypercharge. After symmetry breaking one particle of each doublet becomes
electrically neutral while its partner acquires a positive or a negative
electro-magnetic charge (depending on the choice of the hypercharge).
These particles, like the $AC$ leptons, do not couple to the Higgs boson
and should have masses of the order of $\sim 10^3$ TeV.  Due to
differing self-interaction terms with the photon and the Z-boson the
neutral particle will be slightly lighter than the charged particle and
is therefore the stable state \cite{wells}. Together with its neutrino-like cross section,
the neutral particle constitutes an interesting dark matter candidate \cite{griest}.

\section{Conclusions}

There are two clear-cut achievements of noncommutative geometry in  particle physics:
\begin{itemize}\item
Connes' derivation of the Yang-Mills-Higgs-Dirac-Yukawa ansatz from  the Einstein-Hilbert action,
\item
the fact that this unification of all fundamental forces allows you to  compute correctly the representation content of the Higgs scalar  (i.e. one weak isospin doublet, colour singlet with hyper-charge one  half) from the experimentally measured representation content of the  fermions.
\end{itemize}
The other clear achievements are restrictions on the gauge groups,  severe restrictions on the fermion representations and their  compatibility with experiment.

Finally there are constraints on the top and Higgs masses. They do  rely on the hypothesis of the big desert. Nevertheless we look  forward to the Tevatron and LHC verdict.

\end{document}